\documentclass[a4paper]{ar-1col}
\usepackage[numbers]{natbib}

\usepackage{amsmath}
\usepackage{subcaption}
\newcommand{\bvec}[1]{\boldsymbol {#1}}

\jname{Xxxx. Xxx. Xxx. Xxx.}
\jvol{AA}
\jyear{YYYY}
\doi{10.1146/((please add article doi))}

\begin{document}

\markboth{L. Madeira et al.}{Quantum turbulence in quantum gases}

\title{Quantum turbulence in quantum gases}

\author{L. Madeira,$^1$ M. A. Caracanhas,$^1$ F. E. A. dos Santos,$^2$ and
V. S. Bagnato $^1$
\affil{$^1$Instituto de F\'isica de S\~ao Carlos, Universidade de S\~ao Paulo, S\~ao Carlos, Brazil, 13560-970; email: madeira@ifsc.usp.br}
\affil{$^2$Departamento de F\'isica, Universidade Federal de S\~ao Carlos, S\~ao Carlos, Brazil, 13565-905}
}

\begin{abstract}
Turbulence is characterized by a large number of degrees of freedom,
distributed over several length scales, that result into a disordered
state of a fluid.
The field of quantum turbulence deals with the manifestation of turbulence in quantum fluids, such as liquid helium and ultracold gases.
We review, from both experimental and theoretical points of view,
advances in quantum turbulence focusing on atomic
Bose-Einstein condensates.
We also
explore the similarities and differences
between quantum and classical turbulence.
Lastly, we present challenges and possible directions for the field.
We summarize questions that are being asked in recent works, which
need to be answered in order to understand fundamental properties
of quantum turbulence, and we provide some possible ways of investigating
them.
\end{abstract}

\maketitle


\section{INTRODUCTION}

Turbulence is characterized by a large number of degrees of freedom
interacting nonlinearly, over a substantial range of scales, to
produce a disordered state both in space and time.
Many aspects of classical turbulence (CT) are not well-understood, so
tackling its quantum version, quantum turbulence (QT), seems
ambitious.
It turns out that dealing with turbulence in quantum fluids might be easier
than its classical counterpart, due to the fact that the vortex circulation
is quantized in the former and continuous in the latter.
The advances in trapping, cooling, and tuning the interparticle
interactions in atomic Bose-Einstein condensates (BECs) make them
excellent candidates for studying QT, due to the amount of control
that one can exert over these systems.

\begin{marginnote}[]
\entry{CT}{classical turbulence}
\entry{QT}{quantum turbulence}
\entry{BEC}{Bose-Einstein condensate}
\end{marginnote}

Despite the intrinsic difficulties of this problem, much progress has
been done \cite{tsubota08,white14,tsatos16}.
A milestone was the first observation of turbulence in a trapped BEC
\cite{henn09}, and its signature self-similar expansion.
The occurrence of an energy cascade, demonstrated by the presence of
a power law in the energy spectrum $E\propto k^{-\delta}$, can be considered
as an important step toward the understanding of turbulence
in confined systems \cite{seman11,thompson13,navon16}.

A few words about the scope of this review are in order.
In this work we focus on quantum turbulence in trapped BECs. Comparisons
with turbulence in liquid helium are made when pertinent, but a full review
of the subject is beyond the span of this paper. The same can be said
about classical turbulence.

This review is structured as it follows.
A very brief introduction to quantum gases is given in
Sec.~\ref{sec:quantum_gases}.
In Sec.~\ref{sec:quantum_turbulence} we present the main aspects
of quantum turbulence. We begin with definitions of ``turbulence''
and QT,
Sec.~\ref{sec:nomenclature}.
The picture of a tangle of vortices giving rise to turbulence is
first
introduced in
Sec.~\ref{sec:vortices_turbulence}.
In
Sec.~\ref{sec:introductions}
we contrast aspects of classical and quantum turbulence,
with brief accounts of both.
Some models that have been used to explore QT, namely
the Biot-Savart (BS) and Gross-Pitaevskii (GP) models, are introduced in
Sec.~\ref{sec:theoretical_models}.
Wave turbulence is discussed in
Sec.~\ref{sec:wave_turbulence}.
Section~\ref{sec:vortex_reconnections} deals with one mechanism
that is believed to be essential to QT, vortex reconnections.
Some aspects of two-dimensional turbulence, which is substantially different
from its three-dimensional counterpart, are presented
in Sec.~\ref{sec:2d}.
Section~\ref{sec:misc} contains topics that are not related to the
``usual''
single-component BECs employed in the study of QT, namely bosonic mixtures and fermionic systems.
Section~\ref{sec:experiments}
summarizes the experimental achievements, describing important aspects of
turbulence observed in quantum gases in 3D and 2D.
We emphasize the recent developments in our group, and we show how
the experiments performed in our laboratories relate to the
research on quantum turbulence that is being performed worldwide.
Finally,
in
Sec.~\ref{sec:challenges}
we present some challenges and future issues that,
in our opinion, the field
must face.

\begin{marginnote}[]
\entry{BS}{Biot-Savart}
\entry{GP}{Gross-Pitaevskii}
\end{marginnote}

\section{QUANTUM GASES}
\label{sec:quantum_gases}

\subsection{Bose-Einstein condensates}

Bose-Einstein condensation corresponds to the macroscopic occupation
of the lowest energy quantum state by the particles of a system. This
occurs if the temperature $T$ of the system is cooled below a critical
temperature $T_c$.
Bose-Einstein condensation occurs when the mean interparticle distance
$\bar{l}=(\bar{\rho})^{-1/3}$, $\bar{\rho}$ being the number density of $N$
particles in a volume $V$,
is comparable to the de Broglie wavelength $\lambda_{\rm dB}=h/(mv)$,
where $h$ is Planck's constant, $m$ is the mass of the atoms, and
$v=\sqrt{k_B T/m}$ is their thermal velocity, $k_B$ being
the Boltzmann constant. Imposing $\lambda_{\rm dB}\sim \bar{l}$ implies that a homogeneous gas will undergo a Bose-Einstein condensation at a temperature
$T_c \sim h^2\rho^{2/3}/(m k_B)$.
This simple qualitative argument differs from the accurate result
\cite{pethick08}
only by a factor of $\approx$ 3.3.
The first experimental realizations of Bose-Einstein condensation in
dilute gases were achieved in 1995 \cite{anderson95,bradley95,davis95}, and currently several
laboratories around the world produce BECs on a daily basis.
Most of these experiments are performed on inhomogeneous gases
in harmonic trapping potentials. Then, the critical
temperature \cite{pethick08}
is given
by
$T_c = 0.15 h \bar{\omega} N^{1/3}/k_B$,
where $\bar{\omega}=(\omega_x \omega_y \omega_z)^{1/3}$ is the geometric mean
of the three Cartesian trapping frequencies $\omega_i$, $i=x,y,z$.
Trapping techniques
usually employ magnetic fields or optical means. The
extremely low temperatures
are achieved by laser cooling and evaporation.
One feature of experiments with cold atomic gases that led to rapid advances in the field is the ability to control interactions in the systems. The interatomic interactions and trapping potentials can be changed by modifying external
parameters, such as the applied electromagnetic fields, with
unprecedented control.


\subsection{Superfluid helium}

Liquid helium was first produced in 1908, and 20 years later
scientists noticed that liquid helium had two very distinct phases,
He I and He II, with a singularity in the specific heat, called
$\lambda$-point,
between them \cite{balibar07}.
The behavior of liquid He I was the same as regular liquids, whereas
He II displayed unusual mechanical and thermal properties at low temperatures.
Progress was made with the discovery of superfluidity
\cite{kapitza38,allen38}, flow without viscous dissipation, in direct analogy with the lack of resistance
in a superconductor.

London was the first to connect He II with Bose-Einstein
condensation \cite{london38}, proposing a macroscopic wave function
for the condensed atoms. However,
more attention was given to
the phenomenological two-fluid model introduced by Tisza \cite{tisza38}.
According to this model, liquid He is composed of a normal and a superfluid
part, each of them with its own density and velocity field.
The normal component behaves as a regular fluid, while the superfluid
has no entropy and flows without friction.
Later, Landau formalized the model \cite{balibar17}, which was successful in explaining
a variety of properties and effects in liquid He.
However, a major problem with the two-fluid model is that
it assumes zero superfluid vorticity, the vorticity being the curl of
the velocity field. This is due to the fact that, in the absence of dissipations, the velocity field should be conservative, and thus it can
be written as the gradient of a field, and the curl of a gradient is zero
($\bvec{\omega}=\nabla\times\bvec{v}=\nabla\times\nabla\phi=0$).
Experiments, however, showed clear indications that the vorticity
did not vanish.

\subsection{Quantized vortices}
\label{sec:quantized}

Since Bose-Einstein condensation corresponds to the collective occupation
of the zero momentum state, then a macroscopic wave function can be
used for the $N$ condensed atoms,
$\psi(\bvec{r},t)=\sqrt{\rho(\bvec{r},t)}\exp\left[iS(\bvec{r},t)\right]$,
where $\rho(\bvec{r},t)=|\psi(\bvec{r},t)|^2$ is the condensed density,
$S(\bvec{r},t)$ its phase, and the normalization is given by
$\int d^3\bvec{r} |\psi(\bvec{r},t)|^2=N$.
The probability current is given by
$\bvec{j}=\hbar/(2mi)\left(\psi^*\nabla\psi -\psi\nabla\psi^*\right)=
\rho(\hbar/m)\nabla S$,
where we omitted the spatial and time dependencies hereafter for brevity. This is
a flux of the density $\rho$ that flows with velocity
$\bvec{v}=(\hbar/m)\nabla S$,
$\bvec{j}=\rho\bvec{v}$.
A consequence is that,
when $S$ has continuous
first and second derivatives, the velocity field is irrotational,
$\nabla \times \bvec{v}=0$.
In the presence of a vortex line, a line singularity where $\bvec{v}$
diverges, this does not hold. Thus, a vortex-free velocity
field is irrotational.

The circulation $\Gamma$ around a closed path $C$ is given by
$\Gamma=\oint_C d\bvec{r}\cdot \bvec{v}$.
The macroscopic wave function has to be single-valued, which requires
the phase to change by $2\pi n$, where $n$ is an integer
and it is often called the charge of the vortex, 
when going around the contour. This gives rise to the quanta of circulation
$\kappa=h/m$,
\begin{equation}
\Gamma=\oint_C d\bvec{r}\cdot \bvec{v}=\frac{\hbar}{m} 2\pi n = n \kappa.
\end{equation}
This led Onsager \cite{eyink06} and Feynman \cite{feynman55}
to introduce the quantization of the
vorticity in superfluid helium.
A vortex is an excitation of the system, thus it is a state with higher energy than the
ground-state. It can be shown that the energy is proportional to the
square of the vortex charge, $E \propto n^2$, thus it is energetically
favorable to have $n$ single-charged vortices rather than one $n$-charged
vortex \cite{kawaguchi04}.

\section{QUANTUM TURBULENCE}
\label{sec:quantum_turbulence}

\subsection{Nomenclature}
\label{sec:nomenclature}

The term ``quantum turbulence'' was first used in 1982 by Barenghi in his
Ph.D. thesis \cite{barenghi82}.
Donnelly and Swanson, who adopted the term in 1986 \cite{donnelly86},
were responsible for the shift from the commonly used ``superfluid'' to ``quantum''
turbulence.
This change was not simply a matter of nomenclature, it showed that
turbulence had more to do with the quantization of vortices
than the lack of viscosity of superfluids. Indeed, experiments
later showed that turbulence would decay even in the case of
vanishing viscosity.
The very pertinent question of where the energy goes if there is no friction
was answered by the realization that
vortices decay into sound, as seen in experiments with 2D BECs \cite{kwon14} and numerical simulations using the
Gross-Pitaevskii equation in 2D \cite{stagg15} and 3D \cite{nore97}.

Having justified the term ``quantum turbulence'', we still have to define
``turbulence''. A definition that seems to be accepted is a state
of spatially and temporally disordered flow, with a large number
of degrees of freedom which interact nonlinearly.
Usually, by nonlinear interaction we mean the term $(\bvec{v}\cdot\nabla)\bvec{v}$
in the classical Euler equation,
\begin{equation}
\label{eq:euler}
\frac{\partial\bvec{v}}{\partial t}+(\bvec{v}\cdot\nabla)\bvec{v}
=-\frac{1}{\rho}\nabla p,
\end{equation}
where $p$ is the pressure.
The nonlinear term arises simply from writing Newton's second law for
the continuum.
As we will see, the equation above can be seen as
a particular case of the (incompressible) Navier-Stokes equation
with zero viscosity.

\subsection{From vortices to turbulence}
\label{sec:vortices_turbulence}

Vortex lines can sustain helical deformations, called Kelvin waves,
that move them from a straight orientation. Also, two lines that approach
each other can reconnect and form a cusp, which relaxes into Kelvin
waves.
Disordered combinations of vortex lines can be a manifestation
of quantum turbulence, as proposed by Feynman \cite{feynman55}.
The subsequent work done by Hall and Vinen \cite{hall56_1,hall56_2}
motivated the study of QT using $^4$He.
The recent progress of QT in the context of trapped BECs
came with striking discoveries.
For example, under appropriate conditions, statistical properties
of classical turbulence may arise. One of them is the
Kolmogorov scaling of the energy spectrum ($E\propto k^{-5/3}$),
which suggests an energy cascade from large to small length scales.
This defines what is called Kolmogorov turbulence,
which is found in a specific inertial range as long as there is constant
injection of energy at large length scales.
This requires a self-similar process in which large bundles of
vortices transfer energy to smaller bundles, until the energy reaches
a single vortex.

Numerical simulations showed that if QT obeys the energy spectrum
of the Kolmogorov turbulence, the vortex tangle contains transient
vortex bundles
(vortices with the same orientation) together
with many random vortices
\cite{baggaley12}. These structures are responsible for the large scale,
small $k$, flows.
Although the Kolmogorov scaling has been observed in experiments, these
vortex bundles have not.
Kolmogorov turbulence has been observed in helium at low and high temperatures, $T\ll T_c$ and $T\sim T_c$, respectively. At high
temperatures, the normal fluid dominates, so CT
is expected. However, at low temperatures the situation is more
interesting. Energy transfer occurs because of the non-linear term
$(\bvec{v}\cdot\nabla)\bvec{v}$ in the Euler equation,
Eq.~(\ref{eq:euler}).
This helps to connect both QT and CT.

A different kind of turbulence, called Vinen turbulence,
has been observed in experiments \cite{walmsley08} and
numerically \cite{baggaley12_1}. This kind of turbulence is set
apart from Kolmogorov's because of the absence of large-scale
flow structures.
The question of whether turbulence in trapped BECs is Kolmogorov, Vinen,
a combination of both, or neither, is still unsettled.
However, two vortices in an orthogonal orientation or one doubly-charged
vortex may decay and display a Kolmogorov energy spectrum \cite{zamora15}.
Numerical models using the GP equation also suggest
that trapped BECs should show Kolmogorov turbulence \cite{kobayashi05}.

One last distinction should be made. Stationary turbulence
in 3D (2D) systems requires constant energy injection in large (small)
length scales, which cascades to small (large) scales and is dissipated
with the same rate as injected. In nature, stationary turbulent systems
decay when the energy input stops, characterizing what we
call decaying turbulence.

\subsection{Brief introductions}
\label{sec:introductions}

\subsubsection{Classical versus quantum turbulence}
\label{sec:classical_quantum}

One of the most important differences between classical and quantum turbulence
is the quantization of the circulation, see Sec.~\ref{sec:quantized}.
In this sense, the quantum version of the problem is easier to handle
due to the fact that circulation, which is continuous in the classical
case, can only take a discrete set of values.
Regular fluids are viscous, thus, without a constant energy input,
turbulence decays. At sufficient low temperatures, the thermal component
of the trapped gases is negligible, and it can be considered a pure
superfluid. Even in that scenario, turbulence will decay without constant
energy injection, due to sound waves inside and at the surface of the
condensate.

Trapped BECs also show some interesting features regarding the
amount of control exerted on experiments. The number of atoms,
trapping parameters, and interatomic interactions can be varied to
produce the desired regimes. The drawback is that this freedom
often complicates the comparison between two sets of experiments.

The characteristic length scales are different when
we compare CT and QT and, moreover, the
range available in each case also differs.
In classical fluids, vortices can be as large as the typical length scale
of the system $D$. For a typical BEC the vortex core
is comparable to the healing length $\xi$, and the distance between the vortices
is a few times $\xi$. 

\subsubsection{Classical turbulence}
\label{sec:classical}

The generalization of Euler's equation, Eq.~(\ref{eq:euler}), is
the Navier-Stokes equation. In the case of a solenoidal  incompressible
fluid it can be written as
\begin{eqnarray}
\label{eq:navier_stokes}
\frac{\partial \bvec{v}}{\partial t}+(\bvec{v}\cdot\nabla)\bvec{v}
=
-\frac{1}{\rho}\nabla p+\nu \nabla^2 \bvec{v} + \bvec{g}
\text{ and }
\nabla \cdot \bvec{v}=0,
\end{eqnarray}
where $\bvec{g}$ stands for the external forces.
Reynolds studied the transition from laminar to turbulent
flows in a pipe.
He found that the transition required the Reynolds number $Re=vD/\nu$, 
which
quantifies the intensity of the turbulence
(where $v$ is the flow velocity at a length scale $D$, and $\nu$
is the kinematic viscosity of the fluid), and it can be seen as the ratio of inertial to viscous forces.

The concept of energy cascade came in the 1920s with Lewis Richardson,
who knew that a turbulent steady state required a constant energy injection
at large length scales, and the energy to be dissipated with the
same rate at small length scales. The intermediary range, called
inertial range, was characterized by being independent of the viscosity.
In the 1940s, Kolmogorov formalized the concept of an energy cascade
with a self-similar behavior of the turbulent flow.
In the simplest case of an isotropic and homogeneous steady state,
the energy goes from the largest length scale $D$ to the smallest one
$\eta$, with a constant dissipation rate $\varepsilon$.
The Reynolds number connects these lengths scales,
$\eta/D\approx Re^{-3/4}$.
Kolmogorov suggested that some aspects of turbulence are universal.
Instead of working in real space, it is convenient to go to the
momentum space, to describe the cascade as a function of wavenumbers
$k=2\pi/r$. In the inertial range Kolmogorov showed that the kinetic
energy is given by
\begin{equation}
\label{eq:kolmogorov}
E_k=C\varepsilon^{2/3}k^{-5/3},
\end{equation}
where $C$ is dimensionless and of order one.

Kolmogorov's law, Eq.~(\ref{eq:kolmogorov}), describes three-dimensional
flow. There are only a few examples of physical systems where flow
is truly 2D, in most cases two-dimensional flow is used as an approximation
to a 3D problem, or as building blocks to anisotropic turbulence in 3D.
Nevertheless, 2D classical turbulence is completely different from its
3D counterpart. In 1967 Kraichnan showed that, in 2D, the energy
flows from small to large length scales, in the
opposite direction of the three-dimensional case.
The same exponent, $E\propto k^{-5/3}$, is found, but due to
an inverse energy cascade (IEC). This occurs
simultaneously with a forward cascade of the enstrophy, a quantity
that measures the variance of the vorticity, which makes the energy
spectrum $E \propto k^{-3}$ for large momenta.
\begin{marginnote}[]
\entry{IEC}{inverse energy cascade}
\end{marginnote}

\subsubsection{Quantum turbulence}
\label{sec:quantum}

The visualization of the tangle of vortex lines that constitute turbulence
in a trapped BEC is very difficult, thus much of the progress has been
done relying on numerical simulations. In liquid He experiments,
the determination of the vortex line density $L$ is more
straightforward, and it can be used as a measure of both the
intensity of the QT and the distance between vortices $l\approx L^{-1/2}$.

\begin{figure}[!htb]
\centering
\includegraphics[width=2.75in]{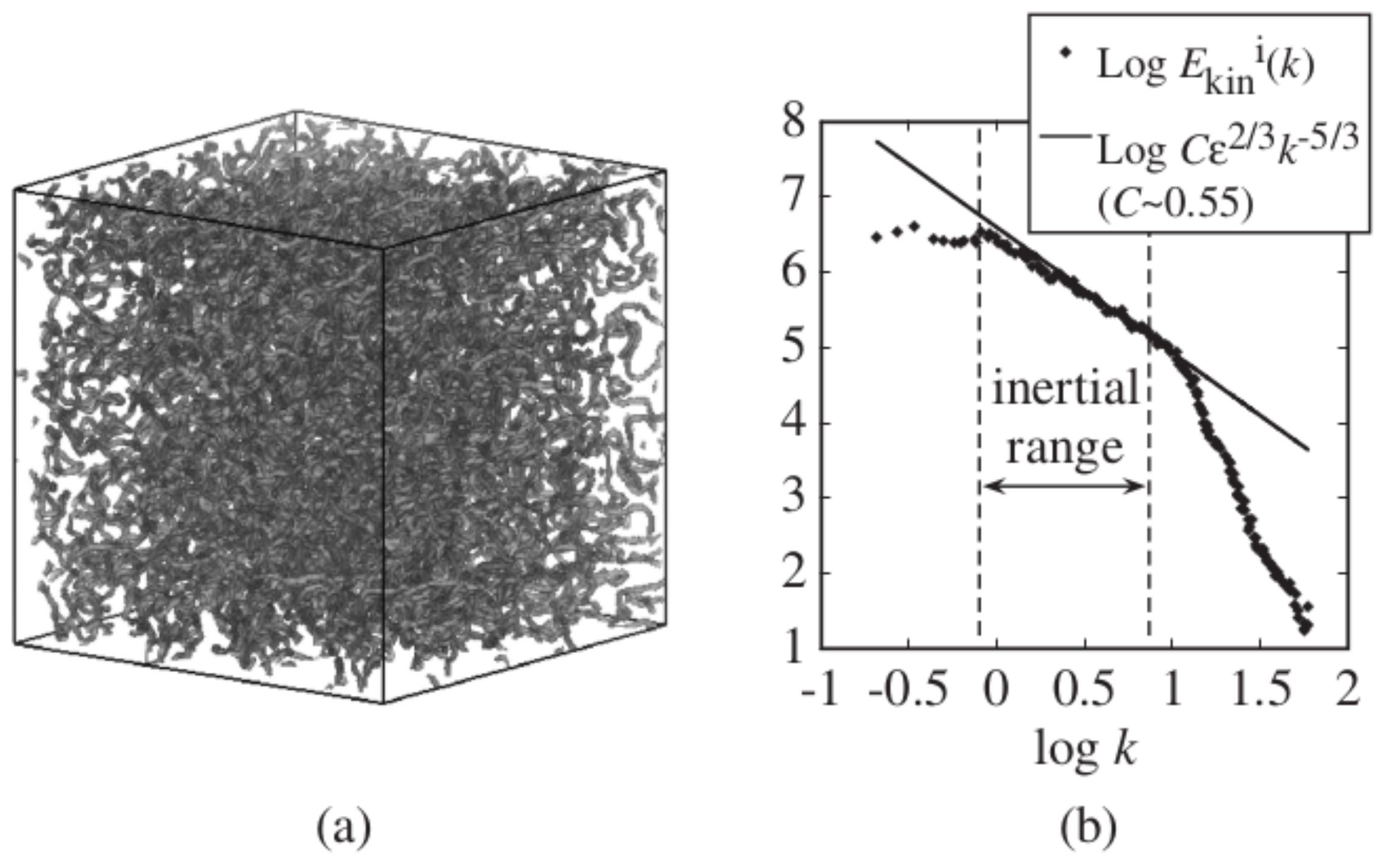}
\caption{(a) Vortex tangle. (b) Spectrum of the incompressible kinetic
energy for the turbulent regime. The points correspond to an ensemble
average of 50 configurations, while the curve stands for the Kolmogorov
law.
Source: Ref.~\cite{tsubota08}.}
\label{fig:tsubota08}
\end{figure}

Not all turbulent states that have been studied correspond to the
Kolmogorov scaling.
They can be of the Vinen kind \cite{volovik03}, but it
also depends on how the excitations are performed.
If the superfluid is thermally driven, then it
lacks energy at the largest scales and it displays
$E\propto k^{-1}$ for large
momenta \cite{baggaley12_3}.
On the other hand, if the superfluid is driven by a turbulent normal
fluid, then the energy spectrum is $E\propto k^{-5/3}$ as is the case
for Kolmogorov turbulence \cite{baggaley12_3}.
Panel (a) of \textbf{Figure \ref{fig:tsubota08}} shows a
vortex tangle in a homogeneous condensate, obtained from
simulations \cite{tsubota08}, while the
Kolmogorov scaling is shown in panel (b).

As we have been arguing throughout this review, one
of the reasons for the
complexity
of QT comes from the large range of length, or conversely momentum, scales available. For the following discussion let us define
$k_l=2\pi/l$ and $k_\xi=2\pi/\xi$, wavenumbers corresponding to the intervortex distance and healing length, respectively.
The discussions, so far, were restricted to the hydrodynamical scale
$k\ll k_l$. Now let us focus on the
range $k_l<k<k_\xi$, $k_\xi$ being the limiting scale because vortices
cannot bend on scales shorter than the vortex core radius, of order of $\xi$.
It is understood that Kelvin waves produce increasingly
shorter wave lengths until the angular frequency of the wave is fast
enough to radiate sound, in a process called Kelvin cascade.
Different theories exist for the energy spectrum of this cascade,
Kozik and Svistunov \cite{kozik04} obtained $E\propto k^{-7/5}$, while
L'vov and Nazarenko \cite{lvov10} predicted $E\propto k^{-5/3}$.
It is still unclear which of the exponents is correct, however
recent studies
\cite{krstulovic12,baggaley14,kondaurova14}
seem to support the spectrum proposed by L'vov and Nazarenko.
Another possibility is that there is a bottleneck between the Kolmogorov
and Kelvin cascades, which changes the shape of the energy
spectrum around $k\approx k_l$ \cite{lvov07}.
This effect is due to the faster rate at which energy flows in the
three-dimensional
Kolmogorov cascade, compared to the one-dimensional Kelvin cascade.
A different approach, but supposing a smooth transition between the
two cascades, has been suggested
\cite{kozik08,lvov08}.


\subsection{Theoretical models}
\label{sec:theoretical_models}

\subsubsection{The Biot-Savart model}
\label{sec:bs}

First introduced by Schwarz \cite{schwarz85}, the Biot-Savart
model parametrizes the vortex line by the curve $s(\zeta,t)$, $\zeta$
being the arclength and $t$ the time. Its name is due to the analogy
with magnetic fields. In this model, the velocity is written as
the curl of a vector potential $\bvec{A}$, $\bvec{v}=\nabla \times \bvec{A}$. The vorticity then obeys the Poisson equation,
$\nabla^2 \bvec{A}=-\bvec{\omega}$, with a solution given by
\begin{equation}
\bvec{A}(\bvec{r})=\frac{1}{4\pi}\int d^3\bvec{r}'\frac{\bvec{\omega}(\bvec{r}')}{|\bvec{r}-\bvec{r}'|},
\end{equation}
for a vortex core located at $\bvec{r}$. The vorticity has
a constant intensity along the vortex core, hence
$\bvec{\omega}(\bvec{r}')d^3\bvec{r}=\kappa d\bvec{\ell}$, which allows
the integration to be performed over a line,
$1/(4\pi)\int d\bvec{\ell} \ \kappa/|\bvec{r}-\bvec{r}'|$.
Notice that $\bvec{A}$ describes an incompressible field,
$\nabla\cdot\bvec{v}=\nabla\cdot\nabla\times\bvec{A}=0$,
however a quantum fluid may have a compressible part which is not
described in this model.
Simulations employing this vector potential
are computationally expensive. Instead, the velocity field may be replaced by
$\bvec{v}_{\rm LIA}=(\kappa/4\pi R) \ln(R/\zeta)
(\bvec{s}'\times\bvec{s}'')$,
where the primes stand for derivatives taken with respect to the arclength, $s'=ds/d\zeta$ \cite{barenghi01}.
This is called local induction approximation due to the fact that nonlocal
contributions to the integral are neglected.
Vortex reconnections, first introduced in Sec.~\ref{sec:vortices_turbulence}, and discussed in Sec.~\ref{sec:vortex_reconnections},
are essential for QT simulations. These are absent
in the BS model and need to be included \textit{ad hoc}
\cite{schwarz85,barenghi01}.

The BS model may be useful for some qualitative behavior of turbulent
quantum fluids, but its assumption of point-like vortex cores, and
the absence of explicit vortex reconnections are severe drawbacks.
In dilute BECs, where during the expansion a single vortex core may be comparable to the
dimension of the whole condensate, this model is not expected to be
successful.

\subsubsection{The Gross-Pitaevskii equation}

The GP approximation was formulated independently by Gross
\cite{gross61}
and
Pitaevskii
\cite{pitaevskii61} in 1961.
The key assumption of this mean-field model
is that all particles are in the same quantum state, corresponding to
zero momentum. Thus the whole system can be described by a
macroscopic wave function $\psi(\textbf{r},t)$, which can be found
by solving 
\begin{equation}
\label{eq:GP}
i\hbar\frac{\partial \psi(\textbf{r},t)}{\partial t}=
\left[
-\frac{\hbar^2}{2m}\nabla^2+V_{\rm trap}+g|\psi(\textbf{r},t)|^2
\right]\psi(\textbf{r},t).
\end{equation}
The parameter $g=4\pi a_s\hbar^2/m$ measures the strength of the
interaction, and it is proportional to the $s$-wave scattering length
$a_s$, while the normalization is given by $\int d^3\bvec{r} |\psi|^2=N$.
Although the GP model is a mean-field approach,
in the limit of a $T=0$ dilute condensate
with two-body repulsive interaction potentials,
the GP equation is exact \cite{lieb00,lieb02}.

Contrary to the BS model, vortex reconnections are present in the solutions, and do not
have to be included \textit{ad hoc}.
The GP equation has been used successfully to
describe a variety of scenarios \cite{dalfovo99}, and
dissipation effects can also be included in it
\cite{tsubota02}.
Finally, we should note that methods beyond a mean-field approach
are available \cite{streltsov06,streltsov07,alon08,wells15}.
A survey on simulations using the GP model, and several other numerical
methods applied to QT, can be found in Ref.~\cite{tsubota17}. 

We can perform a Madelung transformation to $\psi(\textbf{r},t)$ so that
$\psi(\textbf{r},t)=f(\bvec{r},t)\exp
\left[
iS(\textbf{r},t)
\right]$,
where the real functions $f$ and $S$ are associated with the square root of the density and phase of the wave function, respectively.
Substituting
this into Eq.~(\ref{eq:GP}) yields two equations, corresponding to
the real and imaginary components \cite{pethick08},
\begin{eqnarray}
\label{eq:madelung_cont}
\frac{\partial f^2}{\partial t}&=&-\frac{\hbar}{m}\nabla\cdot(f^2\nabla S),\\
-\hbar\frac{\partial S}{\partial t}&=&-\frac{\hbar^2}{2mf}\nabla^2 f
+\frac{1}{2}mv^2+V_{\rm trap}+gf^2,
\label{eq:madelung_2}
\end{eqnarray}
with $v=|\bvec{v}|=|\hbar\nabla S/m|$.
Equation~(\ref{eq:madelung_cont}) is the continuity equation
with $\rho=f^2$ and $\bvec{j}=\hbar f^2 \nabla S/m$, which
shows that the probability $|\psi|^2$ is conserved.
The gradient of Eq.~(\ref{eq:madelung_2}) yields
\begin{equation}
\label{eq:hydro}
\frac{\partial \bvec{v}}{\partial t}=-\frac{1}{m\rho}\nabla p
-\frac{1}{2}\nabla v^2+\frac{1}{m}\nabla\left(
\frac{\hbar^2}{2m\sqrt{\rho}}\nabla^2\sqrt{\rho}
\right)
-\frac{1}{m}\nabla V_{\rm trap},
\end{equation}
where we identified $p=\rho^2g/2$.
Only one term contains $\hbar$, and it is referred to as quantum
pressure, which dominates its classical counterpart only for distances
of the order of (or less than) the healing length. Neglecting
this term yields the dissipation-free Navier-Stokes equation,
Eq.~(\ref{eq:navier_stokes}) with $\nu=0$.
It has been pointed out \cite{santos16} that $S$ is a multivalued field,
thus the chain rule of differentiation cannot be applied to $\exp[iS]$.
Indeed, the curl of Eq.~(\ref{eq:hydro}) would lead us to believe that
vorticity has no dynamics, $\partial \bvec{\omega}/\partial t=0$.
In Ref.~\cite{santos16}, the author derives exact hydrodynamic equations
which present superfluid behavior and include vorticity dynamics.

\subsection{Wave turbulence}
\label{sec:wave_turbulence}

So far we talked about turbulence as a result of the motion and
interaction of vortices. However, some processes in BECs involve
interacting dispersive waves giving rise to turbulence, such as is the
case of sound waves.
We associate the term wave turbulence (WT) to these type of phenomena.
This process also appears as a power law cascade in the energy
spectrum \cite{fujimoto15}.
When the equations of motion describe weakly nonlinear
dispersion, an analytical description is possible \cite{nazarenko11}.
Sound waves are small amplitude excitations to a macroscopic wave
function of the condensate.
Keeping only the smallest nonlinearity \cite{kevrekidis07}
it is possible to analyze plane-wave solutions and their nonlinear
corrections \cite{lvov03}.
The result is a three-wave process which allows the existence
of a steady state characterized by an energy cascade
\cite{nazarenko11,zakharov12}. In the long wavelength regime,
the energy spectrum is of the
Zakharov-Sagdeev type,
$E\propto k^{-3/2}$ \cite{nazarenko11}.
\begin{marginnote}[]
\entry{WT}{wave turbulence}
\end{marginnote}


Wave turbulence may also arise from the vibratory motion of vortex lines,
known as Kelvin waves.
The first theory formulated to explain this type of phenomena involved
a six-wave process \cite{kozik04}.
Later, Nazarenko pointed out that
two different cascades can occur simultaneously in WT, as long as an
even number of waves are involved
\cite{nazarenko11}.
A direct energy cascade flows from large to small length scales,
and there is also an inverse wave action.
Within the Kozik-Svistunov theory \cite{kozik04},
$E\propto k^{-7/5}$ for the direct process, while $E\propto k^{-1}$
for the inverse action.
Kelvin waves occur in classical and quantum fluids \cite{kozik04,nazarenko06}.
In trapped BECs, they have been observed as the result of the
decayment of quadrupole modes 
\cite{bretin03}.
Studies suggests that they may be responsible for the energy transfer
between the scales of intervortex separation down to vortex core
sizes 
\cite{fonda14,kivotides01,tsubota09}.
This feature is present in simulations \cite{kivotides01} using the BS model, see Sec.~\ref{sec:bs}, which does not include a compressible
velocity field. The conclusion is that the cascade enabled by Kelvin
waves involves only vortex energy, independently of phonons or other collective modes.

\subsection{Vortex reconnections}
\label{sec:vortex_reconnections}

Experiments with BECs generate a large number of vortices that reconnect
and tangle with each other
\cite{henn09,white14_2}.
Many experimental techniques for the creation of vortices in dilute BECs
are available
\cite{fetter09,fetter10}.
A vortex reconnection (VR) corresponds to
the approximation of two vortex lines, which then connect and exchange
tails. They are largely responsible for the energy transfer between
different scales, thus VRs are of great importance to the study of QT \cite{serafini17}.
In Ref.~\cite{santos16} it is shown that the creation or annihilation
of a pair of touching vortex lines obeys the power-law $x\propto t^{1/2}$, where $x$ is the distance between the lines and $t$ is the time,
which has also been observed in experiments \cite{paoletti10} and simulations \cite{siggia85}.
In liquid He, VRs have been observed in detail \cite{bewley06}, whereas
the observation in trapped BECs is much more recent
\cite{serafini15}.
Vortex reconnections release kinetic energy and they concentrate vibrations
on individual vortex cores, which in turn can be carried by helical
Kelvin waves.
This is related to an open question in the heart of QT,
the mechanism behind dissipation of $T=0$ frictionless fluids
\cite{fonda14}.
Vinen proposed that high frequency oscillations of a vortex core
can produce phonons in order to dissipate energy in an inviscid fluid
\cite{vinen00}.
\begin{marginnote}[]
\entry{VR}{vortex reconnection}
\end{marginnote}

\subsection{Quantum turbulence in 2D}
\label{sec:2d}

We already mentioned that classical turbulence in 2D is very different from
the 3D case.
Unlike the classical case, where the 2D character is often an approximation
to a 3D problem, truly two-dimensional systems can be realized
with BECs
by exploring the experimental control on the trapping potentials.
This makes dilute cold gases the ideal systems to study two-dimensional
QT (2DQT)
\cite{white14}.
Kelvin waves, the cornerstone for QT decayment in 3D, are not present in
2D, since vortices are zero-dimensional objects in a plane.
\begin{marginnote}[]
\entry{2DQT} {two-dimensional quantum turbulence}
\end{marginnote}

The incompressible kinetic energy spectrum of the 2D turbulent
regime, for a quasiclassical system
with a relatively large $D/\xi$, is given by $E\propto k^{-3}$ for
$k\gg 1/\xi$, and $E\propto k^{-5/3}$ for
$k< 1/\xi$
\cite{bradley12}.
In Sec.~\ref{sec:classical} we stated that the classical
energy spectrum is proportional to $k^{-3}$ for lager momenta
due to a direct
enstrophy cascade. Although in the quantum case the exponent is the
same, the mechanism behind it is completely different.
The $k^{-3}$ scaling is the result of the velocity field profile
of the quantized vortices, which are also responsible for the
enstrophy to be proportional to the number of vortices \cite{white14,bradley12}. The possibility of a vortex-antivortex pair
annihilation makes the enstrophy not to be an inviscid quantity.
This feature of fluctuating number of vortices is highlighted
in GP simulations of Ref.~\cite{numasato10}.
Another study employed simulations of a
purely incompressible fluid and found that
the
IEC takes place only for systems with moderate dissipation
\cite{billam15}.
Further evidence for the IEC includes
dynamical simulations of a forced homogeneous systems
where 
the $k^{-5/3}$ scaling was observed, and also
clusters of vortices with the same sign
\cite{reeves13}.

The first investigations of QT in BECs looked for
quasiclassical characteristics \cite{parker05}.
Unlike the homogeneous case, trapped BECs tend to prevent large scale
motion, so that Vinen turbulence is usually observed.
For example, in Ref.~\cite{white12} the stirring of a BEC, which
was happening at distances of the order of the healing length, was unable
to produce the growth of vortex clustering due to the system size.
When the Kolmogorov scaling can be seen, usually it is for regions
less than a decade \cite{reeves12}, again due to the lack of range of scales available.

Another obstacle to observing QT in 2D is the vortex-antivortex pair
annihilation, which largely prevents vortex clustering.
Experimental protocols for vortex generation end up producing
roughly the same number of positive and negative charge vortices
\cite{kwon14,white14_2,neely10}.
Techniques for successive nucleation of vortices have been proposed 
\cite{sasaki10}.
Results of simulations in a
large, homogeneous BEC clearly show that vortex annihilation is a four-vortex process
\cite{baggaley18}.

\subsection{Miscellanea}
\label{sec:misc}

\subsubsection{Bosonic mixtures}
\label{sec:mixture}

So far we limited ourselves to the discussion of single-component BECs.
However, QT can also be studied in a mixture of two (or more)
bosonic species, with substantially different behavior.
Theoretical investigations of
counterflow turbulence have been carried out employing a mixture
of two bosonic species
\cite{takeuchi10,ishino11}. This would be the analog of QT in $^4$He
at sufficiently high temperatures, when both the normal and superfluid
components are turbulent.
Quantum turbulence can also be studied using condensates with
particles possessing a
spin degree of freedom \cite{mueller06,kurn13}.
The main difference to a regular mixture of bosonic species is
that the population of each spin state is not constant, due to
spin-exchange collisions.
For more details about spin turbulence, the reader is referred to
Ref.~\cite{tsubota14} and references therein.

\subsubsection{Fermionic gases}
\label{sec:fermionic}

Superfluidity can also be achieved in fermionic systems, which
is explained thorough
the Bardeen-Cooper-Schrieffer (BCS) theory of condensation of Cooper pairs
into bosonic-like particles \cite{bardeen57}.
Interest in cold atomic fermionic gases is further augmented by the
BEC-BCS crossover \cite{randeria14}, where the interparticle interactions
can be tuned so that the fermion pairs can change their size from
tightly bound dimers (BEC) to many times the interparticle distance
at the BCS side, passing through the strongly-interacting unitary regime.
A landmark was the production of vortex lattices, throughout the
crossover, in an ultracold $^6$Li gas, demonstrating superfluidity
\cite{zwierlein05}.
The first question that arises is if QT is possible in fermionic gases
and, if so, in which regimes of BEC-BCS crossover
turbulence emerges \cite{bulgac16}.
Apparently, QT is possible in the unitary Fermi gas \cite{wlazlowski15}, and
vortex reconnections were studied in this regime
\cite{wlazlowski15_2}. The microscopic structure of vortices in cold atomic fermionic gases has been studied throughout the BEC-BCS
crossover and in the unitary Fermi gas \cite{mad16,mad17}.
\begin{marginnote}[]
\entry{BCS}{Bardeen-Cooper-Schrieffer}
\end{marginnote}

\subsubsection{Neutron stars}
\label{sec:neutron}

Many problems in nuclear physics are related to quantum turbulence.
Since protons and neutrons are spin-1/2 particles, QT of fermionic
gases is of interest.
In particular, QT may hold the key to a mystery in nuclear astrophysics,
the pulsar glitches. They correspond to sudden increases in the spinning
of neutron stars, while they continually lose angular momentum.
It has been suggested that the
outer core of a neutron star is in a turbulent state, and that
the Reynolds number is related to the glitches
\cite{peralta06,link13}.
The main challenge in this problem
is the disparity between the femtometer scale of vortex cores
and the kilometer scale of neutron stars, however
progress has been made
towards developing a mean field description \cite{khomenko18}.
As is the case with turbulence in trapped BECs, we need
a better understanding of microscopic processes, such as vortex reconnections, that take place in the crust of neutron stars.
Even the study of a single vortex line in neutron matter
is an
active topic of research \cite{madeira19}.

\section{EXPERIMENTS}
\label{sec:experiments}

The study of quantized vortices and QT has increased in intensity with the realization of weakly interacting dilute atomic BECs \cite{anderson95,davis95}. Stimulated largely by the high degree of control which is available within these quantum gases \cite{inouye98,goerlitz01,bloch08,henderson09}, they have been used to investigate quantum turbulence both experimentally and theoretically. The ability to directly resolve the structure of individual vortices and hence the dynamics of a turbulent vortex tangle \cite{bretin03,fonda14,serafini15}, opens the possibility of studying problems which may be relevant to our general understanding of turbulence.

\subsection{The S\~{a}o Carlos Group}

We start by reviewing some of our contributions, both theoretical and experimental, to the field of quantum turbulence, and we show how they are related to research being conducted worldwide.
In 2009, the S\~ao Carlos group was responsible for the first evidences of quantum turbulence in trapped dilute atomic BECs \cite{henn09}.
The experiment consisted of a small vortex tangle created in a harmonically trapped BEC through the combination of rotation and an external oscillating perturbation \cite{henn09,henn09_2}.

\begin{figure}[htp!]%
\centering
\begin{subfigure}[b]{.45\textwidth}
\raisebox{.16\textwidth}{
\includegraphics[width=\linewidth]{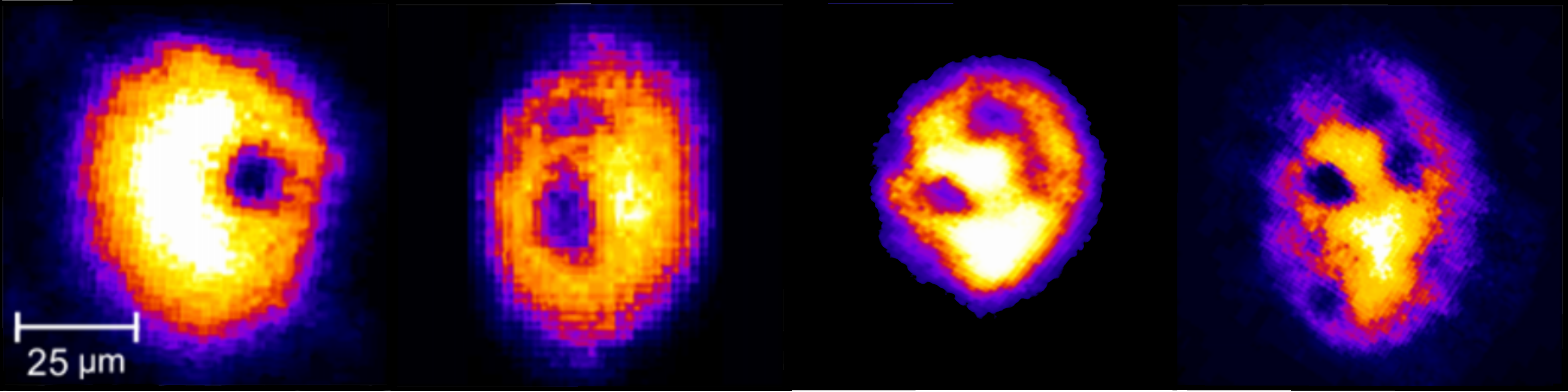}
}
  \caption{Regular vortices}
\end{subfigure}
\begin{subfigure}[b]{.2\textwidth}
\raisebox{.35\textwidth}{
\includegraphics[width=\linewidth]{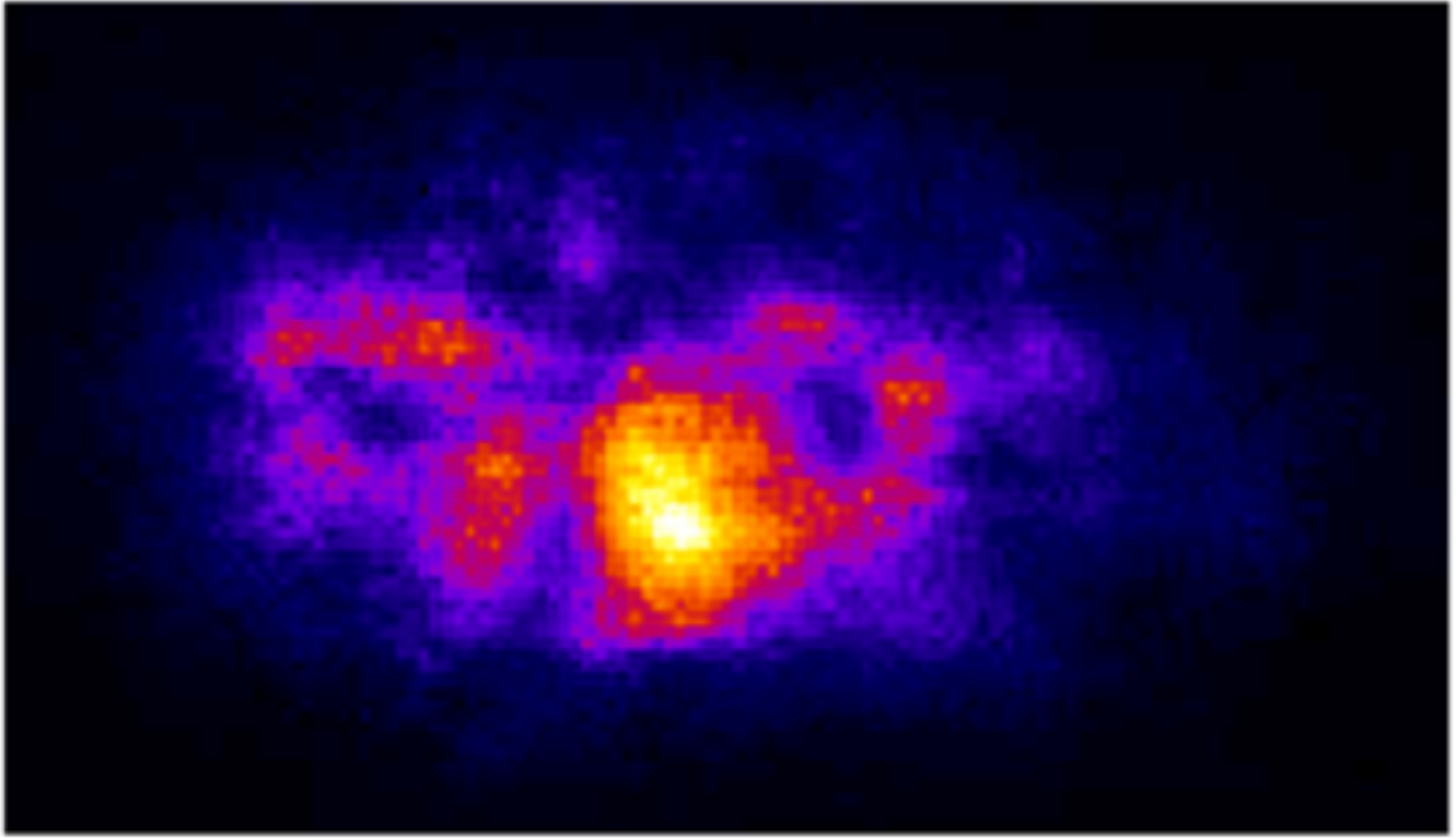}
}
  \caption{Turbulence}
\end{subfigure}
\hspace{0.1cm}
\begin{subfigure}[b]{.20\textwidth}
\includegraphics[width=\linewidth]{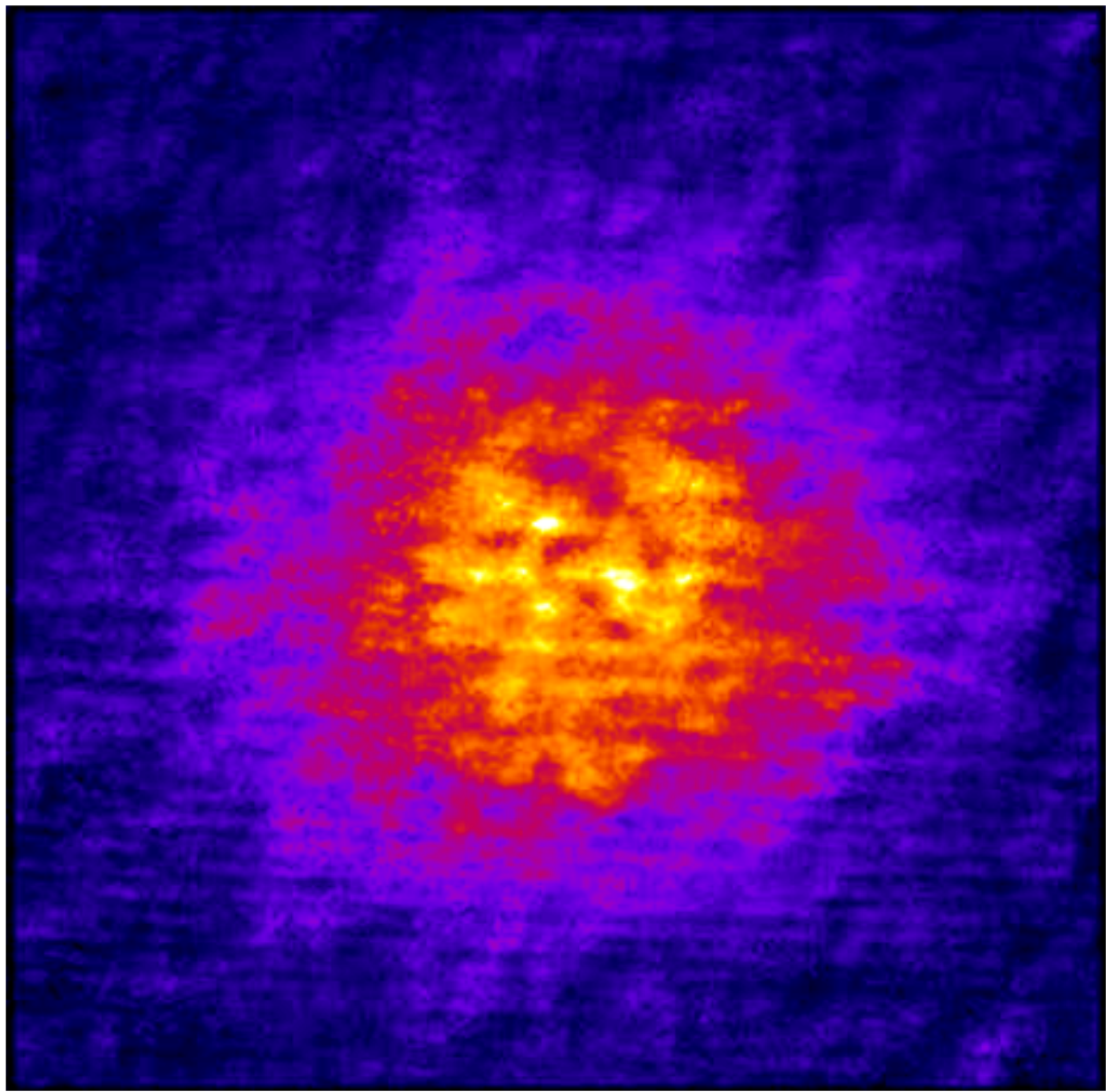}
  \caption{Granulation}
\end{subfigure}%
\vspace{0.5cm}
\caption{Phases of the excited BEC, according to the time and amplitude of the
applied external oscillatory  magnetically field.
From left to right: (a) regular vortices, (b) turbulent cloud, and (c) granulation. Figures extracted from Ref.~\cite{shiozaki13}.}
\label{fig:diagram}
\end{figure}

Different regimes were observed according to the strength and duration of the external oscillatory potential \cite{seman11}, \textbf{Figure \ref{fig:diagram}}. For small amplitude excitations, the only effect was a bending of the main axis of the cloud, irrespective of the duration of the oscillation. Increasing the amplitude of the oscillation caused regular vortices to be nucleated, with the number of vortices increasing monotonically with the application time of the external excitation. Increasing the excitation time further led to a turbulent vortex regime. At very long hold times, the condensate fragmented or granulated, signalizing the decay of the turbulence.

\subsubsection{Finite size system}

Currently, the BEC systems which can be created in the laboratory contain a small number of atoms (a few hundred thousand), and hence do not sustain the number of quantum vortices present in helium experiments. This brings to light the question about the Kolmogorov's scaling, if it still holds in the small range of lengths available.
Despite of the limited number of vortices present in trapped BECs, numerical simulations of these systems \cite{tsubota09,berloff02,kobayashi07,nowak11} suggest that the kinetic energy is distributed over the length scales in agreement with the $k^{-5/3}$ Kolmogorov scaling observed in ordinary turbulence.

Attempts to model
analytically the transition to turbulence in these finite systems
are available in the literature.
The transition from vortices to the turbulent regime
could be established assuming a critical number of vortices, according to the size of the sample \cite{shiozaki11} and to the input energy coming from the external excitation.
In Ref.~\cite{yukalov15}, the authors performed numerical calculations
to reproduce
our experimental conditions. According to the excitation parameters, the perturbed system was classified by a phase diagram. Particular aspects of the granular phase were explored \cite{yukalov14}.

\subsubsection{Self-similar expansion}

The most common diagnostic of trapped atomic clouds is done by imaging them, not in the trap, but after some time of free expansion. That easily distinguishes a thermal from a Bose-condensed cloud. A thermal cloud shows a gaussian density profile that evolves to an isotropic density distribution at long times of expansion. The quantum cloud, in contrast, shows a profile that initially reflects the shape of the confining trap, the Thomas-Fermi regime. In a cigar-shaped trap, for example, the BEC cloud expands faster in the radial than in the axial direction. That causes the signature inversion of the BEC cloud aspect-ratio during the free expansion.

Considering now a turbulent BEC cloud, besides the evidences of the tangle vortices configuration in the density profile (atomic depletion in the absorption image), the cloud free expansion dynamics also differs due to the presence of vorticity. In fact, for the cigar shape trap used in Ref.~\cite{henn09}, the turbulent condensate expands with a nearly constant aspect-ratio once released from its confinement.

To characterize the anomalous expansion of the turbulent sample, a generalized Lagrangian approach was applied in Ref.~\cite{caracanhas13}. The kinetic energy contribution of a tangle vortex configuration was added to the system Lagrangian, and the resulting Euler-Lagrange equations described the dynamics of the cloud.

\subsubsection{Atomic-turbulence and speckle-fields}

Bose-Einstein condensates and atom lasers are examples of coherent matter-wave systems.
On the other hand, an optical speckle pattern can be created by the mutual interference of many light waves of the same frequency, with different amplitudes and phases, giving rise to a random light map.
A very interesting work \cite{tavares17} draws a parallel between a ground-state/turbulent BEC with the propagation of an optical Gaussian beam/elliptical speckle light map.

The researchers analyzed mainly two characteristics of the spatial disorder of the systems. First, measurements of the aspect ratios of regular and turbulent BECs were performed. It is known that for standard BECs there is an inversion of the aspect ratio of the cloud in time-of-flight (TOF) measurements, whereas in the turbulent
case there is a self-similar expansion, without ever inverting its
aspect ratio \cite{caracanhas13}.
For a coherent Gaussian beam,
there is an inversion of the aspect ratio of the waists, whereas it is
preserved in the propagation of the elliptical speckle light map,
just as it is the case with the BEC counterparts.
The second property that was investigated was the coherence in both
systems.
It was found that the correlations in regular BECs resemble the ones in
the Gaussian beam, while the same is true for the turbulent BEC and speckle
beam pair.
This duality opens the possibility of improving our understanding of QT by
looking at statistical atom optics.
\begin{marginnote}[]
\entry{TOF}{Time-of-flight}
\end{marginnote}

\subsubsection{Momentum distribution of a turbulent trapped BEC}

\begin{figure}[!htb]
\centering
\includegraphics[width=2.25in]{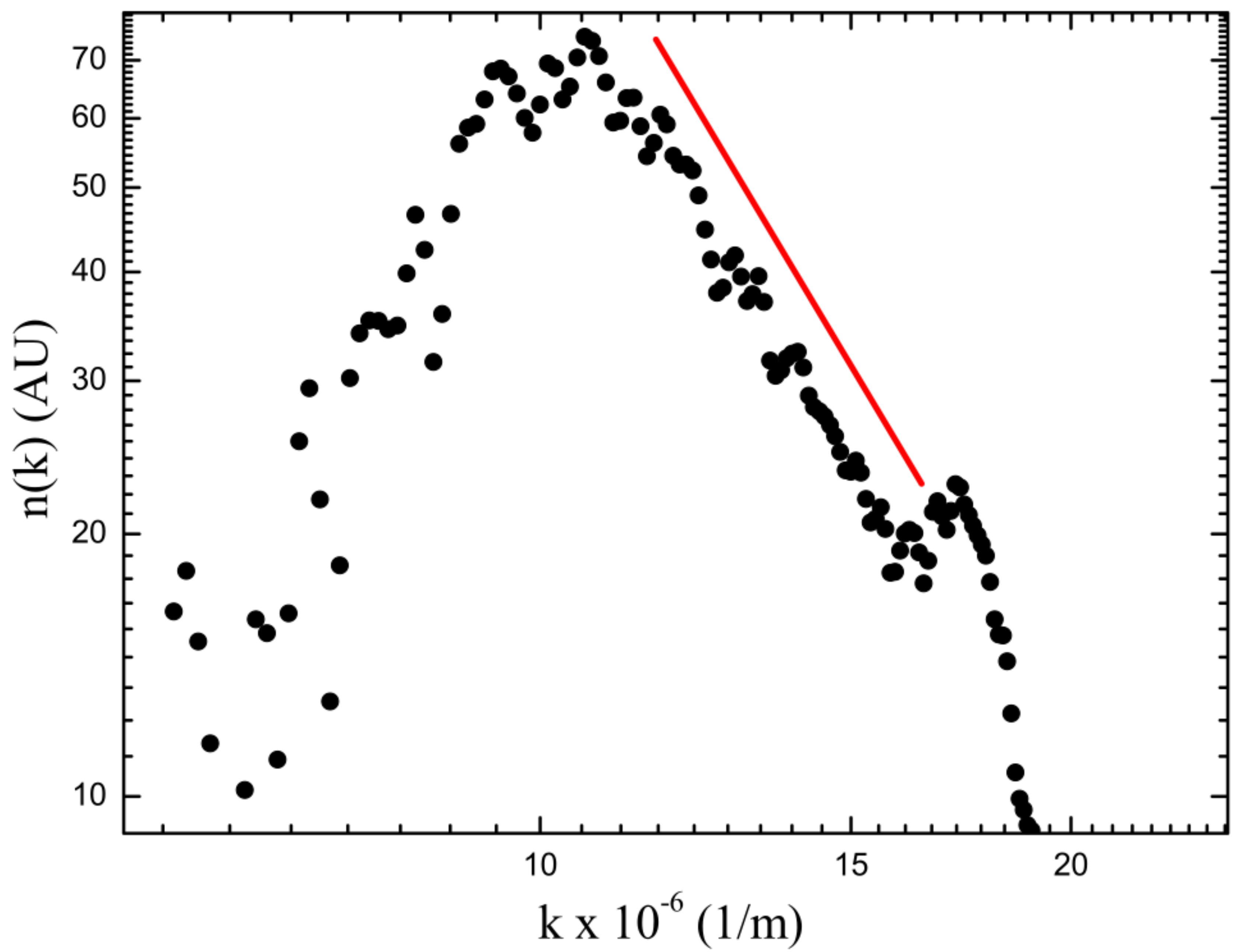}
\caption{ Three-dimensional momentum distribution for a turbulent
cloud
from Ref.~\cite{thompson13}.
The curve corresponds to a line with slope -2.9.}
\label{fig:thompson13}
\end{figure}

To investigate turbulence experimentally, atoms are held for tens of
miliseconds in the trap, after the drive has been terminated, and then they are released for imaging. The atoms are measured by a TOF absorption image technique. In Ref.~\cite{thompson13}, TOF was used to probe the momentum distribution of a turbulent BEC cloud, assuming a ballistic expansion for the atoms.
They found a power-law behavior for the momentum distribution
of $n(k)\propto k^{-2.9}$, \textbf{Figure \ref{fig:thompson13}}.

The momentum distribution is extracted from measurements of the absorption image of the expanded cloud. The projected image is a distorted two-dimensional shadow of the real atomic distribution. As a result, the spatial density has contributions from many wavenumbers along the path of the imaging light. As the interaction is assumed to be negligible, the ballistically expanding atoms allow for an experimental Fourier conversion of the real space density distribution after a TOF to an $in\, situ$ momentum distribution. The radii of the expanded cloud are converted in momentum shells, and the number of atoms are counted in each shell to construct the momentum distribution of the sample.

Remarkably, the turbulent spectra showed a higher-momentum atomic population that is not present in the spectra of a normal BEC. In particular, this unusual region of the spectra showed a distribution which decreases monotonically with the momentum, and its slope in a logarithmic scale can be associated to a power law.

\subsubsection{Collective Modes, Faraday Waves, and Granulation}

\begin{figure}[!htb]
\centering
\includegraphics[width=4.75in]{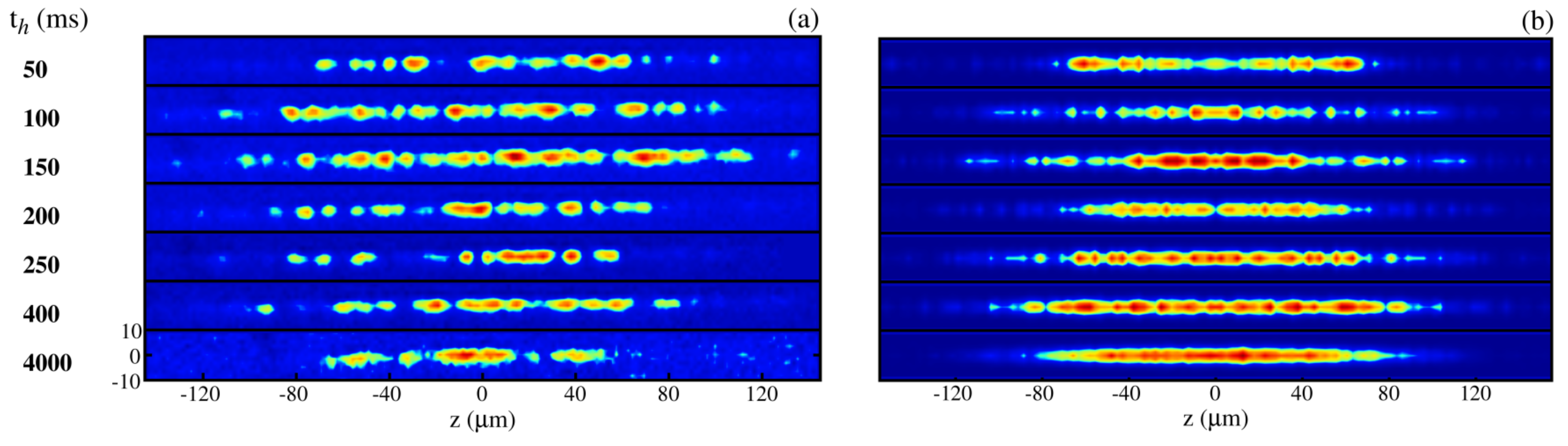}
\caption{Response of an elongated Bose-Einstein condensate to modulated interactions: (a) experiment and (b) numerical simulations.
Granulation is remarkably
persistent in time after the modulation is turned off, and its structure is random between different experimental runs.
Source: Ref.~\cite{nguyen19}.}
\label{fig:faraday}
\end{figure}

In Ref.~\cite{pollack10}, a gas of $^7$Li atoms
was cooled to nearly zero temperature, and the collective modes of an elongated BEC were studied with the modulation of the atomic scattering length. Different regimes appeared by varying the frequency and modulation strength of the external magnetic field. This was explored further, both experimentally and theoretically, in Ref.~\cite{nguyen19}.
Particularly, for modulation frequencies near
twice the trap frequency, longitudinal surface waves are generated resonantly (parametrically).
The dispersion of these waves, also called resonant \cite{nicolin11}
or Faraday waves \cite{nicolin07,engels07}, is well-reproduced by a mean-field theory. Otherwise, far from the resonances (lower modulation frequencies), increasing the modulation strength brings an irregular granulated distribution in the condensate density, \textbf{Figure \ref{fig:faraday}}.
The correlations of this granulated phase could be described with a beyond mean-field theory \cite{lode16}, which characterizes the large quantum fluctuations of this peculiar regime.

\subsection{Two-dimensional quantum turbulence}

Since there is a lack of experimental correspondence, previous numerical studies on 2DQT, which dealt with the energy spectra and vortex dynamics, did not establish any connection with classical 2D turbulence.
Recently, the experiments \cite{neely13,kwon14} showed that even a trapped superfluid sample exhibits many similarities to what is expected from the 2D classical turbulence, extending the universality of 2D turbulence.

To experimentally generate 2DQT, a combined optical-magnetic confinement creates highly oblate BEC, with the turbulent flow resulting from the effective stirring of the center region of the condensate by a repulsive optical potential. This external perturbation induces randomly distributed
nucleation of vortices.
Then the trap is turned off, and the expanded BEC provides the absorption image of the vortices.

In Ref.~\cite{neely13}, the authors investigated, experimentally and numerically, the forced and decaying 2DQT in a BEC. It was demonstrated that the disordered vortex distributions of 2DQT can be sustained, in spite of the vortex-antivortex annihilation. Their observations are indicators of the energy transport from small to large length scales,
corresponding to
the inverse energy cascade.
The thermal relaxation of superfluid turbulence in 2D, based on evidences of the
vortex-antivortex annihilation mechanism, was investigated in Ref.~\cite{kwon14}. They showed the essential role played by vortex-antivortex pairs in 2DQT, characterizing the relaxation of the turbulent condensate through the decay rates of the vortex number.

\section{CHALLENGES AND OPEN QUESTIONS}
\label{sec:challenges}

Trapped superfluid Bose-Einstein condensates show turbulent behavior.
Evidences of energy cascades with power law behavior have been
measured in experiments and supported by theoretical models.
The amount of control that can be exerted on BEC experiments
makes them excellent candidates to investigate quantum
turbulence. Nevertheless, several aspects of this
phenomenon need
experimental investigation and theoretical clarification.
In the following we summarize some of the challenges that
the field must face.

\textbf{Tuning the interactions:}
Throughout this manuscript we pointed out several limitations due to
the range of available length scales in experiments.
A quick estimate of $\log(D/\xi)$,
involving the system size $D$ and the healing length $\xi$,
shows that experiments are limited to one or two decades at best.
However, interactions in trapped BECs can be controlled through
Feshbach resonances \cite{inouye98} to produce smaller healing lengths and
larger cloud radii at the same time, improving the range
of scale lengths available.
How turbulence formation, decay and scaling laws, would react in different interacting environments? Many other questions could be investigated as one studies turbulence in tunable BECs.

\textbf{Finite size:}  Currently, the dilute BECs
which can be created in the laboratory contain a small number of atoms, hence do not sustain the number of quantum vortices present in helium experiments. Fundamental questions exist over the extent to which turbulence can be generated and observed in such small and inhomogeneous systems.

\textbf{Inhomogeneity:} Trapped quantum gases are not
homogeneous systems. What is the influence of this on
characteristics of QT in these systems?
Following the release of the trap, turbulence is expected to be isotropic
after a sufficiently long time. However, experiments are limited to images
a few tens of milliseconds after the trap is turned off.
Models that explicitly take into account differences between the homogeneous
and isotropic behavior of turbulence in bulk fluids and the experimental
conditions of trapped gases, could provide some answers.

\textbf{Probing the turbulent cloud:}
Visualization of the turbulent cloud is of paramount importance for
advances in the field. Techniques for visualizing the vortex
tangle are well-developed in liquid helium systems \cite{bewley06},
and the same level of detail needs to be achieved in trapped condensates.
In this sense, the determination of the kinetic energy spectrum,
and thus the associated turbulence mechanism,
could benefit from
\textit{in situ} measurements of momentum distributions.
The task of extracting the momentum distribution from 2D integrated density profiles of 3D vortex tangles constitutes a huge challenge for
understanding quantum turbulence.

{\bf Kolmogorov's law:} The verification of the $-5/3$ Kolmogorov's law would be by far one of the most spectacular results towards an universal description of turbulence. However, there are questions one has to ask in order to investigate this law in trapped atomic BECs. Does turbulence in BECs achieve the stationary state necessary to the Kolmogorov cascade to be settled? Is the number of vortices in the BEC enough to characterize the Kolmogorov length-scales?

\textbf{Vortex turbulence:}
Questions are being asked about the role played in turbulence by vortex reconnections, and about the nonclassical quantum turbulent regime (called ``Vinen'' or ``ultraquantum'' turbulence), which  is different from ordinary turbulence in terms of energy spectrum and decay.
The wide range of length scales available in turbulence
gives rise to energy cascades with power-law behavior. Measuring the
exponent does not mean that the mechanism behind it is understood,
specially when models predict exponents spaced closely together.
Kolmogorov turbulence assumes vortex bundles, however, due to the
finite size of trapped BECs, their formation is unlikely to happen.

{\bf Mechanism of vortex tangle generation:}
Theoretical simulations point that a certain degree of dissipation has to be introduced in the Gross-Pitaevskii equation in order to generate vortices, the phenomenological dissipative part representing the effects of the thermal cloud \cite{kobayashi07,proukakis08,white10,baggaley11}.
In previous works, there are
evidences of counter-flow in excited Bose-Einstein condensates, with thermal and condensate components moving out-of-phase and against each other \cite{tavares13}. The presence of vortices in the interface between the BEC and the thermal cloud has been observed, indicating that the thermal cloud has indeed some degree of contribution in the vortex nucleation.
In that sense, the study of vortex generation and evolution to a vortex tangle as a function of the size of the thermal component must
be investigated.

\textbf{Generation and decay of turbulence:} What are the most effective and efficient ways to generate
turbulence? Does the way in which the turbulence is generated affect the ``type'' of turbulence
created?
Also, one important aspect to characterize in the experiments is how turbulence decays. To observe this process, one needs high-resolution and non-destructive imaging of the trapped cloud.
How can one quantify this decay experimentally, i.e.,  what are the experimental observables that allow to quantify decay rates?

\textbf{``Exotic'' systems:}
Systems that depart from the standard single-component BEC, that is usually
employed in experiments and simulations, may shed light on
aspects of QT that have not been explored yet.
Recently, turbulence in a dipolar Bose gas has been achieved \cite{bland18},
opening the possibility to study QT with long range interactions.
Another example consists of bosonic mixtures of two
\cite{myatt97,modugno02,papp08,thalhammer08,mccarron11,pasquiou13,
wacker15,barbut14,delehaye15,yao16,desalvo17,roy17,laurent17,wu18}, or possibly
more, species that could be used in QT experiments. Besides the intrinsic
differences between the behavior of the two systems, one species
could be used as means to visualize the other. Another
possibility for multi-component gases is using BECs with spin degrees
of freedom.
Much of this review was based on bosonic fluids, however
fermionic gases can also display superfluidity. The study of
QT in cold atomic Fermi gases is still incipient, and much remains
to be done.


\textbf{Change in the dimensionality:} A highly oblate, but still 3D, BEC may suppress superfluid flow along the tight-confining direction, providing 2D superfluid vortex dynamics. Hence, 2D turbulence can be obtained \cite{rooney11}. These systems open the possibility of investigating transitions between QT in two and three dimensions.

\textbf{Beyond mean-field:} We should also note that much of the theoretical framework relies on mean-field theories.
Effects beyond mean-field need to be investigated, either
to show that the Gross-Pitaevskii model is valid, or pinpoint
its limitations.



%

\section*{DISCLOSURE STATEMENT}
The authors are not aware of any affiliations, memberships, funding, or financial holdings that
might be perceived as affecting the objectivity of this review. 

\section*{ACKNOWLEDGMENTS}
We thank
M.C. Tsatsos, P.E.S. Tavares, A. Cidrim, A.R. Fritsch, and
C.F. Barenghi
for the useful discussions.
This work was supported by
the S\~ao Paulo Research Foundation (FAPESP)
under the grants 2018/09191-7 and 2013/07276-1.
We also thank Centro de Pesquisa em \'Otica e Fot\^onica (CePOF) for
their
financial support.
F.E.A.S. acknowledges CNPq for support through Bolsa
de Produtividade em Pesquisa No. 305586/2017-3.

\bibliographystyle{ar-style4}
\bibliography{ref.bib}

%



%
%
%
%
%
%
%

\end{document}